\documentstyle[twoside,fleqn,espcrc2,epsf]{article}

\newcommand{\AmS}{{\protect\the\textfont2
  A\kern-.1667em\lower.5ex\hbox{M}\kern-.125emS}}

\title{Some remarks on Abelian dominance\thanks{Research supported 
in part by U.S.\ Department of Energy grant
DE-FG02-92ER-40672, OTKA Hungarian Science Foundation T 017311 and
the Physics Research Group of the Hungarian Academy of Sciences,
Debrecen.}}

\author{Tam\'as G.\ Kov\'acs\address{Department of Physics,
                University of Colorado,\\
                Boulder CO 80309-390, USA}
        and 
        Zsolt Schram\address{Department of Theoretical Physics, 
                Kossuth Lajos University,\\
                Debrecen H-4010, Hungary}}

\begin{document}
\begin{abstract}
We used a renormalisation group based smoothing to address two questions
related to Abelian dominance. Smoothing 
enabled us to extract the
Abelian heavy-quark potential from time-like Wilson loops 
on Polyakov gauge projected  configurations. We obtained a very 
small string tension which is inconsistent with the string tension 
extracted from Polyakov loop correlators. This shows that the Polyakov 
gauge projected Abelian configurations do not have a consistent 
physical meaning. We also applied the smoothing on SU(2) configurations 
to test how sensitive Abelian dominance in the maximal Abelian gauge 
is to the short distance fluctuations. We found that on smoothed 
SU(2) configurations the Abelian string tension was about 30\% smaller 
than the SU(2) string tension which was unaffected by smoothing. This
suggests that the approximate Abelian dominance found with the Wilson 
action is probably an accident and it has no fundamental physical 
relevance.
\end{abstract}

\maketitle

\section{INTRODUCTION}

It is an old idea to try to understand non-Abelian gauge theories
in terms of an effective Abelian model with a smaller symmetry
group. One possible way of doing this on the lattice is to isolate
$U(1)^{N-1}$ link variables belonging to a maximal torus of SU(N).
This is called Abelian projection.
The hope is that non-Abelian confinement might be explained as a 
condensation of monopoles in the resulting Abelian projected model
(see e.g.\ \cite{Polikarpov} for a recent review).
If one wants to explain the non-Abelian physics in the Abelian 
projected system, a necessary condition is that the Abelian model
has to reproduce the physical features of the non-Abelian system.
This property is referred to as Abelian dominance.

The projection procedure necessarily involves some gauge fixing. 
In principle the physical properties of
the projected system can depend on the gauge choice. 
Up to now the only gauge in which the Abelian 
projected system seems to capture the physics of the non-Abelian 
model is the maximal Abelian gauge \cite{MAG}.
Here in the SU(2) case the Abelian and non-Abelian 
string tensions at Wilson $\beta=2.51$ agree to within 8\% \cite{Bornyakov}. 
In other gauges, most notably in the Polyakov gauge (where Polyakov 
loops are diagonalised) the situation is more controversial. 
Since all the Polyakov loops can be exactly diagonalised at the 
same time, in this case ``Abelian dominance'' exactly and trivially
holds if the string tension is measured with Polyakov loop correlators.
On the other hand due to the high level of noise on the projected 
configurations, it is impossible to extract the string tension 
from Wilson loops \cite{Suzuki}. 

In this talk we discuss some related issues.
The first question we address is that of the gauge 
choice. We use a recently proposed smoothing technique based on
renormalisation group ideas \cite{DeGrand}. We can drastically 
reduce the short-distance fluctuations while preserving the 
long-distance physical properties of our configurations, most
importantly the SU(2) string tension. This allows us to extract 
the heavy quark potential from Wilson loops on Polyakov gauge projected 
configurations. The resulting Abelian string
tension turns out to be practically zero. This result is inconsistent
with the string tension measured from Polyakov loop correlators.
It shows that the physical meaning of Polyakov gauge projected 
configurations is questionable. 

The only gauge known to us in which approximate
Abelian dominance has been found (with the Wilson action) is the maximal
Abelian one. Therefore in the second part of the talk we shall 
concentrate only on this gauge.
We study the question, how Abelian
dominance depends on the details of the short-distance fluctuations
in this particular gauge. 
Using the above mentioned smoothing on Monte Carlo 
generated SU(2) gauge configurations
we can produce smoothed configurations with the same long-distance
properties but reduced short-distance fluctuations. Comparing the
Abelian string tension on the original and the smoothed configurations
we can gain insight into its dependence on the short-distance details.
For a more detailed account of this work the reader is referred to Ref.\ 
\cite{KS}.

\section{THE GAUGE CHOICE}

The very idea of Abelian 
dominance is that the diagonal Abelian degrees of freedom can account
for the physical properties of the full non-Abelian configurations.
The issue of gauge fixing is definitely important here since the
part of the system that we retain/discard with the Abelian projection
very strongly depends on it. 

Let us consider the Polyakov gauge first.
On any given SU(2) configuration all
the links belonging to the Polyakov loops can be diagonalised 
simultaneously by a suitable gauge transformation. Therefore any 
physical quantity derived from Polyakov loops will be trivially 
and exactly reproduced after Abelian projection in this gauge. 
In particular there is exact Abelian dominance for the string 
tension measured with Polyakov loop correlators \cite{Ejiri}.

A good test of whether the Polyakov gauge projected Abelian 
configurations capture some genuine physics would be to 
measure the string tension using time-like Wilson loops and
compare this to the string tension obtained with Polyakov loop
correlators. Unfortunately this cannot be done directly because
the gauge fixing introduces so much noise that one would need
a huge number of configurations to get enough statistics. 

We can however use an ensemble of smoothed configurations
and do all the measurements on them. 
We generated an ensemble of 20 $12^4$ configurations with the
fixed point action of Ref.\ \cite{DeGrand} at $\beta=1.5$ which
corresponds to a physical lattice spacing of 0.144 fm. After one
smoothing step we measured both the full SU(2) and 
the Polyakov gauge projected U(1) heavy quark potential on them
using time-like Wilson loops. We used the method and computer code
of Heller et al.\ \cite{Heller}. 
\begin{figure}[htb]
\vspace{-0.3cm}
\begin{minipage}{7.5cm}
\epsfxsize=7cm \epsfysize=7cm
\epsfbox{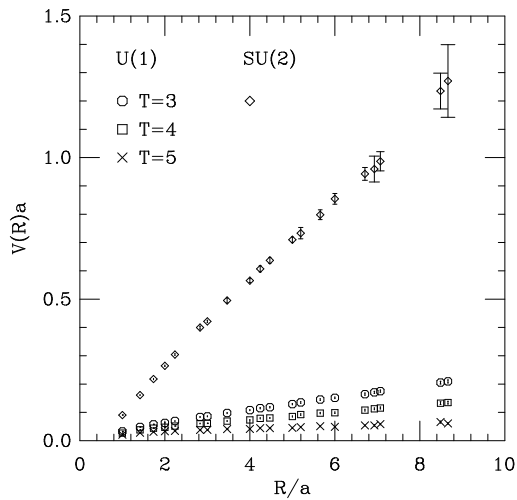}
\end{minipage}
\vspace{-1cm}
\caption{
The static quark potential measured with timelike Wilson
loops. Diamonds correspond to the full SU(2) potential, the other 
three symbols represent the U(1) potential measured in the Polyakov 
gauge with Wilson loops of different time extensions.}
\label{fig:pot_pg}
\vspace{-0.5cm}
\end{figure}
Our results are shown in Figure \ref{fig:pot_pg}.
In the SU(2) case we have a good plateau at $T=3$ (this has also 
been confirmed on another ensemble of larger statistics) but in the
U(1) case the potential decreases considerably with increasing 
$T$ even at this point. One can
conclude that in the $T \rightarrow \infty$ limit the U(1)
string tension is probably very close to zero. 

The discrepancy is striking. We would also
like to note that the static quark potential measured by Polyakov-loop
correlators is exactly the
same as the full non-Abelian potential. We also note that the string 
tension obtained from Polyakov loop correlators and timelike Wilson loops
should be the same (up to some small finite size effects).
This means that two different  but  physically
equivalent measurements of the same physical quantity give
absolutely different results on the Polyakov gauge projected 
configurations.
Our result for the Polyakov gauge strongly suggests that
the physics of the Abelian projection is not only very strongly
gauge dependent but in most of the arbitrarily chosen gauges 
the Abelian projected configurations do not even have a consistent 
physical meaning.

The maximal Abelian gauge (MAG) is special as it minimises the off-diagonal 
components of the link degrees of freedom, the ones that
are discarded in the projection \cite{MAG}. For this reason 
the MAG is a priori a better choice than the gauges that diagonalise
an arbitrarily selected set of operators like the Polyakov loops.

\section{ABELIAN DOMINANCE AND SHORT RANGE FLUCTUATIONS}

In this section we study how Abelian dominance in the
maximal Abelian gauge depends on the precise nature of 
short distance fluctuations.

We generated 100 $8^3 \times 12$ lattices with the fixed 
point action of Ref.\ \cite{DeGrand} at $\beta=1.5$ (lattice 
spacing $a=0.144$ fm). At first as a check we verified that 
Abelian dominance holds for this ensemble. We transformed 
the configurations into the maximal Abelian gauge. 
This was done using the usual overrelaxation 
procedure iterated until the change in the gauge fixing action
became less than $10^{-8}$ per link. After Abelian projecting these 
configurations the heavy quark potential was extracted from 
time-like Wilson loops in the same way as in the previous section.
From the heavy-quark potential 
we obtained $\sigma_{na}=0.123(7)$ for the non-Abelian and 
$\sigma_{ab}=0.119(5)$ for the Abelian string tension in lattice 
units.

After this check we applied one step of smoothing to the same 
ensemble of SU(2) configuration and repeated the measurement of
the Abelian and non-Abelian potential on the smoothed 
configurations. It gave
$\sigma_{na}=0.115(9)$ and $\sigma_{ab}=0.080(10)$ for the
SU(2) and the U(1) string tension respectively.

The SU(2) string tension on the smoothed configurations is essentially
the same as on the unsmoothed ones, reflecting the fact that smoothing
does not change the long-distance features. On the other hand, 
as a result of smoothing, the Abelian string tension dropped by 
about 30\%. This shows that the Abelian string tension is very 
sensitive to the details of the short-distance fluctuations on 
the SU(2) configurations. A similar result has been found for the
monopole string tension using cooling with the Wilson action 
\cite{Hart}. 

It seems to us  
quite impossible to reconcile this fact with the expectation that the Abelian 
string tension is a genuine long-distance physical observable 
which is in some sense equivalent to the SU(2) string tension. 
In view of this, the approximate Abelian dominance found with Wilson 
action in the maximal Abelian gauge seems to be an accident rather than
a fundamental physical phenomenon.

\end{document}